\documentclass[conference]{IEEEtran}
\IEEEoverridecommandlockouts

\usepackage{cite}
\usepackage{amsmath,amssymb,amsfonts, mathtools, array}

\usepackage{graphicx}
\usepackage{textcomp}
\usepackage{xcolor}
\usepackage{algorithm}
\usepackage{algpseudocode}
\usepackage{caption}
\usepackage{subcaption}
\usepackage{tabularx}
\usepackage{float}
\usepackage{lipsum,mathtools}
\usepackage{multirow}
\usepackage{dirtytalk}
\usepackage[numbers]{natbib}
\def\BibTeX{{\rm B\kern-.05em{\sc i\kern-.025em b}\kern-.08em
    T\kern-.1667em\lower.7ex\hbox{E}\kern-.125emX}}
    \IEEEaftertitletext{\vspace{-2\baselineskip}}
    \usepackage[left=0.57in,right=0.57in,top=0.7in,bottom=0.95in]{geometry}
\begin{document}
\title{Multimodal Multiple Federated Feature Construction Method for IoT Environments\\
}
\author{\IEEEauthorblockN{Afsaneh Mahanipour}
\IEEEauthorblockA{\textit{Department of Computer Science} \\
\textit{University of Kentucky}\\
Lexington, KY, USA \\
ama654@uky.edu}
\and
\IEEEauthorblockN{Hana Khamfroush}
\IEEEauthorblockA{\textit{Department of Computer Science} \\
\textit{University of Kentucky}\\
Lexington, KY, USA \\
khamfroush@cs.uky.edu}
}
\maketitle
\begin{abstract}
The fast development of Internet-of-Things (IoT) devices and applications has led to vast data collection, potentially containing irrelevant, noisy, or redundant features that degrade learning model performance. These collected data can be processed on either end-user devices (clients) or edge/cloud server. Feature construction is a pre-processing technique that can generate discriminative features and reveal hidden relationships between original features within a dataset, leading to improved performance and reduced computational complexity of learning models. Moreover, the communication cost between clients and edge/cloud server can be minimized in situations where a dataset needs to be transmitted for further processing. In this paper, the first federated feature construction (FFC) method called multimodal multiple FFC (MMFFC) is proposed by using multimodal optimization and gravitational search programming algorithm. This is a collaborative method for constructing multiple high-level features without sharing clients' datasets to enhance the trade-off between accuracy of the trained model and overall communication cost of the system, while also reducing computational complexity of the learning model. We analyze and compare the accuracy-cost trade-off of two scenarios, namely, 1) MMFFC federated learning (FL), using vanilla FL with pre-processed datasets on clients and 2) MMFFC centralized learning, transferring pre-processed datasets to an edge server and using centralized learning model. The results on three datasets for the first scenario and eight datasets for the second one demonstrate that the proposed method can reduce the size of datasets for about 60\(\%\), thereby reducing communication cost and improving accuracy of the learning models tested on almost all datasets.
\end{abstract}


\section{Introduction}
The fast development of Information and Communication Technologies (ICTs) has led to the invention and development of numerous smart devices and applications, including Internet-of-Things (IoT) devices such as smartphones and autonomous vehicles that generate a huge volume of data. The collected data are fed into different machine learning (ML) models to learn patterns from the data and extract knowledge about the environment. They were traditionally sent to the a cloud server, but because of real-time response requirements and privacy issues, for many IoT applications the data cannot be sent to the cloud and need to be processed either locally or on the edge \cite{nishio2019client}. These data are high dimensional big data that may contain some irrelevant, redundant, noisy, or heterogeneous ones that can negatively impact execution time and performance of the ML models. Also it may increase the communication cost between end-user devices and edge servers. Therefore, in IoT environments, data pre-processing methods can be applied on local datasets of each device to reduce data size and consequently, improve the model performance and communication cost \cite{zebari2020comprehensive}.

Among data pre-processing methods, feature selection (FS) and feature construction (FC) are two common approaches in data mining and ML. FS methods select informative features from the original feature set. FC methods construct high-level features by combining informative ones with suitable operators to extract hidden relationships among original features. Some FC methods use evolutionary computation algorithms that are divided into two main categories: single-feature construction and multiple-feature construction. These methods converge to one best solution that can either contain one or multiple new features \cite{tran2019genetic}. However, different feature combinations may lead to the similar learning performance. Therefore, FC problem can be investigated from multimodal optimization (MO) perspective. MO techniques like niching techniques try to find multiple global optimum solutions by controlling population diversity and dividing the population into multiple sub-populations \cite{yang2016adaptive}. 

To construct new features in IoT environments with distributed datasets, a collaborative FC method is needed. There are a number of federated FS (FFS) methods in the literature that are inspired by federated learning (FL) procedure to select informative features in a collaborative manner \cite{cassara2022federated}, \cite{feng2022vertical}. However, according to our best knowledge, FC has not been investigated from this perspective. Also, FC has not been investigated as a problem with multimodal property. In this paper, we propose to use multimodal niching Gravitational Search Programming (GSP) algorithm for federated feature construction (FFC). Among niching strategies, the crowding strategy is adopted to construct multiple new features by dividing the population according to the distance between each program and a random reference point. The proposed method reduces data size in each client in an effective manner without information loss. This leads to speeding-up ML models, and reducing their complexity especially in privacy-preserving learning scenarios, and communication cost reduction when datasets need to be sent to edge servers. The main novelties of the proposed method can be summarized as follows:
\vspace{-1mm}
\begin{itemize}
\item Proposing the first FFC method inspired by FL procedure
\item Adopting a MO technique for solving FC problem for the first time
\item Constructing multiple features by using a filter-based FC approach in one execution
\item Comparing the proposed method with other centralized FC methods, and other FFS methods in the literature, resulting in a mean accuracy increase of about 4.5\(\%\) and 1\(\%\), respectively while reducing data size for about 60\(\%\) 
\end{itemize}


\section{BACKGROUND AND RELATED WORKS}
Previous works on FS and FC have been mainly done centralized. To the best of our knowledge, there has been limited research on federated and distributed FS, and no research on FFC. In the following, we will explain the previous works more in depth.

\subsection{Centralized Feature Construction}
The quality of the feature set is a crucial factor in the performance of supervised machine learning techniques such as classification. To enhance performance, two main data pre-processing approaches are commonly used: FS and FC. FC methods can be classified into three categories: filter, wrapper, and hybrid methods. Our paper proposes a filter-based FC method. Unlike wrapper methods, filter methods do not rely on learning algorithms and FC is treated as a pre-processing step that does not interact with the learning algorithms. These methods evaluate and rank constructed features based on inherent characteristics such as information theory, mutual information, or correlation criteria \cite{ma2020filter}. 

FC is considered as NP-hard problem because it is not practical to find the exact solution with large amount of features. Genetic programming (GP) is mostly used for constructing either one or multiple new features \cite{koza1994genetic}. GP uses complex tree representation and complicated process with crossover and mutation. Single FC methods cannot provide a good accuracy by using one constructed feature. On the other hand, most of multiple FC methods use multi-tree representation for each program that contains a number of trees correspond to constructed features. This may cause disappearing good constructed features in programs with low fitness during the evolutionary process. Also, some methods execute single FC method for multiple times to construct multiple features. To the best of our knowledge, FC has not been investigated in IoT environments with distributed local datasets. To address these challenges, we propose using GSP algorithm for FFC and adopting a multimodal niching method for solving FC problem and constructing multiple features. 
\vspace{-1mm}
\subsection{Federated Feature Selection (FFS)}
There are a few number of FFS methods in the literature that are divided into two main categories: vertical FFS \cite{li2023fedsdg} and horizontal FFS \cite{hu2022multi}. In vertical FFS, datasets in clients have instances with the same IDs, and different feature sets. However, in horizontal FFS, clients have different instances with the same feature sets. The horizontal FFS in \cite{cassara2022federated} used filter-based Cross-Entropy FS method in both clients and edge server. A few number of informative features were selected with information loss. Therefore, the proposed method cannot provide a good trade-off between communication cost and accuracy. Another work \cite{mahanipour2023wrapper} proposed a wrapper-based horizontal FFS method. Gravitational search algorithm (GSA) was used as a local and global feature selector in clients and edge server, respectively. The results demonstrated that by selecting about half number of features, the method could achieve a reasonable trade-off between communication cost and accuracy. 

Some methods use a trusted third party on edge server instead of a global FS algorithm. Therefore, there is not a global algorithm to guide the FFS process. In \cite{hu2022multi}, a horizontal FFS method based on particle swarm optimization (PSO) was proposed. Binary bare-bones PSO selected features from local datasets in clients and a trusted third party assembled the selected features. This may lead to a non-accurate FFS since there is not a global FS algorithm on the server to continue evolutionary process. Another work \cite{zhang2023federated} proposed an unsupervised horizontal FFS. Initially, a feature average relevance one-class support vector machine algorithm was employed to eliminate irrelevant features by combining the idea of outlier detection and dimension reduction. Then, a feature relevance hierarchical clustering algorithm was used to group related features and create distinct feature clusters. In this method, the FS procedure was executed only once on clients' datasets unlike the concept of FFS. After reviewing existing methods and their challenges, rather than solely selecting features, our objective is to construct multiple distinctive features to achieve a better balance between accuracy and cost. To accomplish this, we incorporate MO with GSP algorithm inspired by horizontal FFS.
\vspace{-1.5mm}
\subsection{Gravitational Search Programming (GSP)}
\vspace{-1mm}
GSP is one of the latest automatic programming approaches that has been proposed in 2019 \cite{mahanipour2019gsp}. This algorithm is a population-based meta-heuristic algorithm that used GSA as a process to construct mathematical expressions, automatically, and it is suitable for FC. By contrast to other programming methods, GSP used a fixed-length string representation for encoding programs instead of complex tree structure with great depth. Terminal and internal nodes of programs can be selected from problem variables and predefined mathematical operators, respectively. This algorithm is faster and simpler than others as it eliminated genetic operators including crossover, mutation, selection and replacement in its process. The GSP search procedure can be narrated by several main steps:

\begin{enumerate}
  \item Initializing a random population of programs using variables and operators for terminal and internal nodes.
  \item Iterating the following sub-steps until reaching a stopping criterion:
  \begin{enumerate}
    \item Evaluation: calculating fitness value of each program by an appropriate fitness function.
    \item Updating the best and the worst programs.
    \item Calculating mass and acceleration of each program using GSA formulas.
    \item Updating the velocity and the position of programs using GSA formulas.
  \end{enumerate}
  \item Returning the best program with the maximum fitness value as the optimum solution.
\end{enumerate}
You can read more details about this algorithm in \cite{mahanipour2019gsp}.

\section{Proposed Method}
\vspace{-1mm}
In this section, the architecture and details of the proposed filter-based multimodal multiple horizontal FFC is presented. 

\subsection{System Overview}
We consider a 2-tier system, where horizontal FL is performed. A number of clients like IoT devices are considered as the first-tier that collect data, and a single edge server is considered as the second-tier (This approach is generalizable to more number of edge servers). We consider that there are a set of \(M\) clients \(C_m\), \(m:=\{1, 2, ..., M\}\) and an edge server \(e\). Here, \(M\) should be at least 2 because if we have only one client then the problem will be a centralized FC. The set of all local datasets hold by the clients is represented by \(D=\{D_1, ..., D_M\}\). Each client's dataset \(D_m=\{X, Y\}:=\{(x_n, y_n)\}_{n=1}^N\), containing \(N\) unique samples and the same feature set \(F=\{f_1, f_2, ..., f_L\}\). \(x_n=(x_{n1}, x_{n2}, ..., x_{nL})\) is a sample vector, and \(Y=(y_{1}, y_{2}, ..., y_{N})\) is the class label vector of \(N\) samples, where \(y_n\in \{1, ..., ClassLabel\}\). \(F=\{f_1, f_2, ..., f_L\}\) is the original feature set of each client that is used to select informative ones and combine them to construct multiple more discriminative features \(\{f_{q1}\}\), \(\{f_{q2}\}\), ..., \(\{f_{q\beta}\}\).
\vspace{-1mm}
\subsection{Proposed Algorithm}
As mentioned in previous sections, FFC is considered as a MO problem, and niching strategy is applied to solve it. The proposed MMFFC method is composed of two phases the same as FL procedure: local and global phase. The GSP algorithm is used for FC in both phases as feature constructor. To construct multiple features, the local GSP population is partitioned into sub-populations in each client and leverages local data to obtain the optimal constructed feature for its respective region. Then the global GSP in edge server aggregates the constructed features by clients and continues the FC process. The global and local feature constructors communicate iteratively until reaching the stopping criterion. The pseudo code of global and local feature constructors on edge server side and client side are given in Algorithm 2 and 3, respectively. Fig. 1 illustrates the FFC procedure as well.

\textbf{Local Phase at Clients:} In this phase, local feature constructor investigates original features to select and combine them with appropriate mathematical operators to construct high-level discriminative features (Algorithm 3). For this purpose, the local GSP algorithm for FC starts with randomized population of programs (Line 1 Algorithm 3). Each program in the population corresponds to a constructed feature. A fixed-length string represents the structure of each program, and its dimension is specified by the number of original features in the dataset (\(L\)), the depth of programs (\(PD\))and the number of operands (\(NO\)) for internal nodes of programs. This dimension is calculated as follow:
\vspace{-4mm}
\begin{equation}\label{my_first_eqn}
Program'sDimension = L+\sum_{i=1}^{\ PD} NO^i+\sum_{j=0}^{\ (PD-1)} NO^j
\end{equation}

\noindent The string contains binary bits (for selecting informative features from original feature set) and integer bits (for selecting operators of internal nodes and connection links). At the next step, crowding clustering niching method is applied to the GSP algorithm to maintain its population diversity and find multiple solutions (Line 3 Algorithm 3). The population of programs is divided into multiple sub-populations by calculating the distance between each program and the random reference point by using hamming distance. Similar programs with small differences are grouped in the same cluster,and different programs are separated into different clusters (Lines 1-13 Algorithm 1). This distance metric provides the number of different bits between two program strings. Crowding clustering niching pseudo code is given in Algorithm 1.

Then, each program in each niche/cluster is evaluated by a fitness function and a fitness value is assigned to it (Line 5 Algorithm 3). Here, information gain ratio (IGR) is used as a fast filter-based fitness function to measure the effectiveness of a constructed feature. The IGR of a constructed feature \(f_{q\beta}\) is indicated by \(IGR\)(\(f_{q\beta}\)), and the information gain (IG) of the \(f_{q\beta}\) is indicated by \(IG\)(\(f_{q\beta}\)). \(H\)(\(f_{q\beta}\)) indicates the entropy of a constructed feature \(f_{q\beta}\). Then, the \(IGR\)(\(f_{q\beta}\)) is calculated as follow:
\vspace{-2mm}
\begin{equation}\label{my_third_eqn}
IGR(f_{q\beta})=\frac{IG(f_{q\beta})}{H(f_{q\beta})}
\end{equation}

\noindent The IG of the constructed feature is the difference between the \(H(Y)\), information entropy of the class labels \(Y\), and \(H(Y|f_{q\beta})\), conditional entropy of the class labels given the constructed feature. \(IG\), \(H(Y)\), \(H\)(\(f_{q\beta}\)), and \(H(Y|f_{q\beta})\) are defined in Eq. (3), (4), (5), and (6), respectively.

\vspace{-1mm}
\begin{equation}\label{my_third_eqn}
IG(f_{q\beta})=H(Y)-H(Y|f_{q\beta})
\end{equation}
\vspace{-4mm}
\begin{equation}\label{my_forth_eqn}
H(Y)=-\sum_{i=1}^c P(Y_i)log_2 P(Y_i)
\end{equation}

\noindent where \(c\) is the number of class labels, and \(P(Y_i)\) is the probability of the class label \(Y_i\) in the training set.
\vspace{-1mm}
\begin{equation}\label{my_fifth_eqn}
H(f_{q\beta})=-\sum_{j=1}^N P(V_{q\beta} ^j) log_2 P(V_{q\beta} ^j)
\end{equation}

\noindent where \(N\) is the number of values of the \(f_{q\beta}\). \(V_{q\beta} ^j\) is the \(j\)-th value of the constructed feature \(f_{q\beta}\).
\vspace{-1mm}
\begin{equation}\label{my_sixth_eqn}
H(Y|f_{q\beta})=-\sum_{j=1}^N P(V_{q\beta} ^j) \sum_{i=1}^c P(Y_i|V_{q\beta} ^j)log_2 P(Y_i|V_{q\beta} ^j)
\end{equation}

\noindent where \(P(Y_i|V_{q\beta} ^j)\) is the conditional probability of the \(i\)-th class given the \(j\)-th value of the constructed feature \(f_{q\beta}\).

In this method, as the population is divided into multiple sub-populations, the mass, position, velocity, and acceleration of programs within each niche are updated based on the fitness values of the niche population, including the best and worst programs found in the current iteration (Lines 6-8 Algorithm 3). After reaching the stopping criterion, maximum iteration here, clients save the last updated programs' positions, velocities, and their corresponding fitness values to be used in the next iteration (Line 10 Algorithm 3). 

\textbf{Global Phase at Edge Server:} The number of programs in each niche or niche size \((NS)\) is randomly selected in each iteration for all clients similar to what has been done in \cite{hu2021multimodal}. Therefore, the number of niches \((A)\) is calculated as: \(A=ceil (S/NS)\) where \((S)\) is the number of programs in the population, and \(ceil\) represents the ceiling of the value (Line 3 Algorithm 2). Multiple programs with the highest fitness value that construct multiple high-level discriminative features can be specified by using MO method in each client after reaching the stopping criterion. Therefore, clients need to save all these programs with their corresponding fitness values and their indices in the local GSP population. All these programs are used as the population of the global GSP (Lines 4-8 Algorithm 2). Then, by using the GSP updating formula, the programs are updated and returned to their corresponding clients (Lines 9-12 Algorithm 2). This communication between clients and edge server will be continued until reaching the stopping criteria.
\vspace{-2mm}
\begin{algorithm}
\caption{Pseudo code of the Crowding clustering method}\label{alg:cap}
\begin{algorithmic}[1]
\renewcommand{\algorithmicrequire}{\textbf{Input:}}
\renewcommand{\algorithmicensure}{\textbf{Output:}}
\Require The number of programs in the population (\(S\)), Number of niches (\(A\)), and niche size (\(NS\))
\Ensure Desired number of niches
\newline
\State Initialize a reference point \(R\) randomly in the population
\(s\) $\leftarrow$ \(S\)  
\hfill// \(s\) is the number of unclustered programs
\For {\(i=1:A\)}
    \If{$s>NS$} 
        \State $ns\leftarrow NS$
    \Else
        \State $ns\leftarrow s$
     \hfill// \(ns\) is the niche size for the particular niche
    \EndIf 
    \State Find the closest program \(Z\) to \(R\) in population
    \State Fine \(ns-1\) closest programs to \(Z\) in population
    \State Put \(Z\) and \(ns-1\) programs in the \(i\)-th niche
    \State Remove the \(ns\) selected programs from the population
    \State \(s\) $\leftarrow$ \(s-ns\)
\EndFor
\end{algorithmic}
\end{algorithm}
\vspace{-3mm}

\begin{algorithm}
\caption{Pseudo code of the global phase at edge server}\label{alg:cap}
\begin{algorithmic}[1]
\renewcommand{\algorithmicrequire}{\textbf{Input:}}
\renewcommand{\algorithmicensure}{\textbf{Output:}}
\Require Multiple local \say{Best} programs, their corresponding fitness values, and their indices in the local population from all clients
\Ensure Multiple \say{Best} solutions (Best constructed features) 
\newline
\While{reaching the stopping criteria or maximum iteration}
\For {all clients}
    \State Select \(NS\) randomly and compute \(A\)
    \State Execute Algorithm 3
    \State Receive multiple best programs, their corresponding fitness values and indices
\EndFor
\State Global population = all local best programs
\State Global fitness = corresponding fitness values of local best programs
\State Updating Kbest, G, Best, and M based on the GSP
\State Calculating global programs' acceleration and velocity
\State Updating global programs' position 
\State Send the updated programs to their clients
\EndWhile
\end{algorithmic}
\end{algorithm}
\vspace{-1mm}

\begin{algorithm}
\caption{Pseudo code of the local phase at client}\label{alg:cap}
\begin{algorithmic}[1]
\renewcommand{\algorithmicrequire}{\textbf{Input:}}
\renewcommand{\algorithmicensure}{\textbf{Output:}}
\Require The number of programs, Number of features, Depth of programs, Number of operands, \(A\), and the updated \say{Best} programs of the client from the global algorithm
\Ensure Multiple local \say{Best} programs and their corresponding fitness values, and indices
\newline
\State\textbf{Initialization:} Initial a population of fixed-length strings randomly
\State In iterations \(>\) 1, use saved population from last iteration and replace the updated \say{Best} programs

\State Use Algorithm 1 to cluster the population
\While{reaching the stopping criteria/maximum iteration}
\State Evaluate programs by IGR (2)
\State Updating Kbest, G, Best, and M for each niche (GSP)
\State Calculating the acceleration and velocity of each niche population (GSP)
\State Updating programs' position of each niche (GSP)
\EndWhile
\State save the position, velocity, and fitness of local programs
\end{algorithmic}                    
\end{algorithm}
\vspace{-3mm}
\section{Experimental Results}

In this section, the proposed method is evaluated through two scenarios. 1) MMFFC federated learning involves applying MMFFC to the local datasets to decrease their sizes and using vanilla FL as a learning model. 2) MMFFC centralized learning involves transferring reduced-size datasets to an edge server after applying the MMFFC to the clients' datasets and then feeding them to a centralized learning model.

\subsection{Datasets}

According to the literature, three benchmark datasets are used for the first scenario that represent IoT network characteristics. These datasets are: IoT network traffic (DEFT) dataset for device classification, ACC dataset for anonymized credit card transactions to predict fraud transactions, and KDD99 dataset for network intrusion detection. The characteristics of these datasets are provided in Table I. For the second scenario, eight benchmark datasets are collected from the University of California, Irvine (UCI) machine learning repository. The details of these datasets are demonstrated in Table II.

\begin{table}[htbp]
\vspace{-1mm}
\caption{Characteristics of the IoT datasets}
\vspace{-4mm}
\begin{center}
\begin{tabular}{|c|c|c|c|}
\hline
\textbf{Dataset name} & \textbf{\textit{Classes}}& \textbf{\textit{Features}}& \textbf{\textit{Instances}} \\
\hline
\textbf{DEFT}& 16 & 111 & 7289\\
\hline
\textbf{ACC}& 2 & 30 & 284807\\
\hline
\textbf{KDD99}& 5 & 41 & 494021\\
\hline
\end{tabular}
\label{tab1}
\end{center}
\vspace{-5mm}
\end{table}
\vspace{-2mm}

\begin{table}[htbp]

\caption{Details of the UCI datasets}
\vspace{-4mm}
\begin{center}
\begin{tabular}{|c|c|c|c|}
\hline
\textbf{Dataset name} & \textbf{\textit{Classes}}& \textbf{\textit{Features}}& \textbf{\textit{Instances}} \\
\hline
\textbf{Wine}& 3 & 13 & 178\\
\hline
\textbf{Sonar}& 2 & 60 & 208\\
\hline
\textbf{wdbc}& 2 & 30 & 569\\
\hline
\textbf{HillValley}& 2 & 100 & 606\\
\hline
\textbf{Ionosphere}& 2 & 34 & 351\\
\hline
\textbf{Balance-Scale}& 3 & 4 & 625\\
\hline
\textbf{Iris}& 3 & 4 & 150\\
\hline
\textbf{Thyroid}& 3 & 5 & 215\\
\hline
\end{tabular}
\label{tab1}
\end{center}
\vspace{-5mm}
\end{table}
\vspace{-1mm}

\subsection{Evaluation Measure}
\vspace{-1mm}
In this work, vanilla FL algorithm (FedAvg) with multi layer perceptron (MLP) with 6 layers is used in the first scenario. In the second one, after constructing new features, the reduced-size datasets are transferred and fed to a centralized C4.5 decision tree classifier. Moreover, classification accuracy (acc) and feature reduction (fr) as measures of learning model performance and communication cost are used. These two metrics are calculated as follow: \(acc = \frac{C}{T}*100\) where \(C\) is the number of correctly classified instances in the test set, and \(T\) is the total number of instances in the test set. The higher the \(acc\), the constructed features are more distinctive. \(fr = \frac{TF - CF}{TF}*100\) where \(TF\) is the total number of features in the original set, and \(CF\) is the number of constructed features. The higher the \(fr\), the communication cost is lower, and the FC method is more appropriate.
\vspace{-2mm}
\subsection{Parameter setting}
In our experiment, a FL with 10 clients is considered. Independent and identically distributed (iid) and non-iid configurations are used for data distribution. Therefore, the results are comparable with results in \cite{cassara2022federated}, \cite{hu2022multi}, \cite{mahanipour2023wrapper}, and \cite{zhang2023federated}. The number of programs and iterations in local GSP are 30 and 5, respectively. The total number of iterations in global GSP is 100, and total best programs from all clients form its population. Other parameters of GSP are unchanged and the same as the original algorithm\cite{mahanipour2019gsp}. These parameters are obtained by trial and error.
\vspace{-1mm}
\subsection{Results and Analysis}
\vspace{-1mm}
In the first scenario, as the proposed method is the first method in federated feature construction, it is compared with the baseline with no FS (No-FS) and four existing FFS methods in the literature on three benchmark datasets: DEFT, ACC, and KDD99. However, in the second scenario, the proposed MMFFC method is compared with two recent centralized FC methods on eight UCI datasets. The results are demonstrated in Table III and IV. In the first scenario, filter-based FFS method proposed in \cite{cassara2022federated} (Fed-FS-CE) can select a few number of features and reduce communication cost, but lost a lot of information. However, our method can construct multiple high-level features and achieve good accuracy with low computational complexity of local learning models. For example, for ACC dataset, although Fed-FS-CE \cite{cassara2022federated} and MFPSO \cite{hu2022multi} methods remove 80 \(\%\) and 56 \(\%\) of features respectively, they cannot provide accurate results. Moreover, FSHFL \cite{zhang2023federated} and Fed-FS-GSA \cite{mahanipour2023wrapper} methods obtain the same or about the same accuracy of the FL classifier with no feature selection (No-FS). However, the proposed method can simultaneously enhance accuracy by approximately 1.6\(\%\) and decrease data size by 60\(\%\). The results for both iid and non-iid datasets show that the MMFFC can enhance mean accuracy by 1\(\%\) and decrease data size by about 65\(\%\). In the second scenario, instead of sending the entire dataset, only the constructed features are sent and fed into the C4.5 classifier at the edge server. We observed that the proposed MMFFC method can achieve higher accuracy for nearly all datasets while reducing the average communication cost by approximately 57\(\%\). However, for one or two datasets, it can obtain the same accuracy with more number of constructed features. For example, for Balance-Scale dataset the proposed method with 3 features and Fcm \cite{ma2019hybrid} with 10 features can obtain almost the same accuracy, and FCMFS \cite{ma2020filter} obtains better result with fewer number of features and communication cost. 

\begin{table*}[htbp]
\caption{Comparison between the proposed method and No-FS and four state-of-the-art FFS methods in the literature}

\begin{center}
\begin{tabular}{|c|c|c|c|c|c|c|c|c|c|}
\hline
Dataset & Method & (\#) Features & FR (\%) & CA (\%) & Dataset & Method & (\#) Features & FR (\%) & CA (\%) \\ \hline
\multirow{6}{*}{DEFT (iid)} & No-FS & 111 & - & 95.25 & \multirow{6}{*}{DEFT (non-iid)} & No-FS & 111 & - & 94.73\\
 & MMFFC & 21 & 81.08 & \textbf{96.13} & & MMFFC & 17 & 84.68 & \textbf{94.83}\\
 & Fed-FS-GSA \cite{mahanipour2023wrapper} & 63 & 43.24 & 94.91 & & Fed-FS-GSA \cite{mahanipour2023wrapper} & 57 & 48.64 & 93.55 \\
 & Fed-FS-CE \cite{cassara2022federated} & 11 & 90.09 & 29.57 & & Fed-FS-CE \cite{cassara2022federated} & 9 & 91.89 & 28.02 \\
 & MFPSO \cite{hu2022multi} & 48 & 55.48 & 82.79 & & MFPSO \cite{hu2022multi} & 49 & 55.85 & 82.38 \\
 & FSHFL \cite{zhang2023federated} & 87 & 21.62 & 95.38 & & FSHFL \cite{zhang2023federated} & 84 & 24.32 & 94.46 \\
 \hline
\multirow{6}{*}{ACC (iid)} & No-FS & 30 & - & 96.41 & \multirow{6}{*}{ACC (non-iid)} & No-FS & 30 & - & 94.59\\
 & MMFFC & 12 & 60.00 & \textbf{98.02} & & MMFFC & 9 & 70.00 & \textbf{96.84} \\
 & Fed-FS-GSA \cite{mahanipour2023wrapper} & 17 & 43.33 & 95.39 & & Fed-FS-GSA \cite{mahanipour2023wrapper} & 21 & 30.00 & 93.11\\
 & Fed-FS-CE \cite{cassara2022federated} & 6 & 80.00 & 37.68 & & Fed-FS-CE \cite{cassara2022federated} & 5 & 83.33 & 33.61 \\
 & MFPSO \cite{hu2022multi} & 13 & 56.66 & 80.65 & & MFPSO \cite{hu2022multi} & 12 & 60.00 & 79.24 \\
 & FSHFL \cite{zhang2023federated} & 26 & 13.33 & 96.41 & & FSHFL \cite{zhang2023federated} & 26 & 13.33 & 95.60 \\\hline
\multirow{6}{*}{KDD99 (iid)} & No-FS & 41 & - & 99.74 & \multirow{6}{*}{KDD99 (non-iid)} & No-FS & 41 & - & 99.50\\
 & MMFFC & 23 & 43.9 & 99.01 & & MMFFC & 22 & 46.34 & 98.32 \\
 & Fed-FS-GSA \cite{mahanipour2023wrapper} & 25 & 39.02 & 97.14 & & Fed-FS-GSA \cite{mahanipour2023wrapper} & 26 & 36.58 & 96.93 \\
 & Fed-FS-CE \cite{cassara2022federated} & 11 & 73.17 & 24.36 & & Fed-FS-CE \cite{cassara2022federated} & 8 & 80.48 & 21.08\\
 & MFPSO \cite{hu2022multi} & 20 & 51.21 & 92.52 & & MFPSO \cite{hu2022multi} & 18 & 56.09 & 92.19\\
 & FSHFL \cite{zhang2023federated} & 20 & 51.21 & \textbf{99.82}& & FSHFL \cite{zhang2023federated} & 20 & 51.21 & \textbf{99.64} \\\hline
\end{tabular}
\label{tab1}
\end{center}
\vspace{-5mm}
\end{table*}

\begin{table}[htbp]
\vspace{-3mm}
\caption{Comparison between the proposed method and two existing centralized FC methods in the literature}
\vspace{-2mm}
\begin{center}
\begin{tabular}{|c|c|c|c|c|}
\hline
Dataset & Method & (\#) Features & FR (\%) & CA (\%) \\ \hline
\multirow{5}{*}{Wine} & No-FS & 13 & - & 83.96 \\
 & MMFFC & 4 & 69.23 & \textbf{97.36} \\
 & Fcm \cite{ma2019hybrid} & 10 & 23.07 & 92.78 \\
 & FCMFS \cite{ma2020filter} & 4.3 & 66.92 & 91.94 \\
 \hline
\multirow{5}{*}{Sonar} & No-FS & 60 & - & 68.72 \\
 & MMFFC & 10 & 83.33 & \textbf{92.45} \\
 & Fcm \cite{ma2019hybrid} & 10 & 83.33 & 71.72 \\
 & FCMFS \cite{ma2020filter} & 5.5 & 90.83 & 72.61 \\
 \hline
\multirow{5}{*}{wdbc} & No-FS & 30 & - & 92.22 \\
 & MMFFC & 8 & 73.33 & \textbf{97.72} \\
 & Fcm \cite{ma2019hybrid} & 10 & 66.66 & 95.62 \\
 & FCMFS \cite{ma2020filter} & 5.03 & 83.23 & 95.75 \\\hline
 \multirow{5}{*}{HillValley} & No-FS & 100 & - & 50.38 \\
 & MMFFC & 23 & 77.00 & \textbf{99.55} \\
 & Fcm \cite{ma2019hybrid} & 10 & 90.00 & 99.33 \\
 & FCMFS \cite{ma2020filter} & 3.28 & 96.72 & 99.00 \\\hline
 \multirow{5}{*}{Ionosphere} & No-FS & 34 & - & 86.27 \\
 & MMFFC & 12 & 64.70 & \textbf{92.38} \\
 & Fcm \cite{ma2019hybrid} & 10 & 70.58 & 90.65 \\
 & FCMFS \cite{ma2020filter} & 4.60 & 86.47 & 89.54 \\\hline
 \multirow{5}{*}{Balance-Scale} & No-FS & 4 & - & 77.07 \\
 & MMFFC & 3 & 25.00 & 98.39 \\
 & Fcm \cite{ma2019hybrid} & 10 & - & 98.67 \\
 & FCMFS \cite{ma2020filter} & 2.10 & 47.5 & \textbf{99.26} \\\hline
 \multirow{5}{*}{Iris} & No-FS & 4 & - & 93.48 \\
 & MMFFC & 3 & 25.00 & \textbf{96.67} \\
 & Fcm \cite{ma2019hybrid} & 10 & - & 93.11 \\
 & FCMFS \cite{ma2020filter} & 4.63 & - & 92.30 \\\hline
 \multirow{5}{*}{Thyroid} & No-FS & 5 & - & 89.37 \\
 & MMFFC & 3 & 40.00 & \textbf{95.56} \\
 & Fcm \cite{ma2019hybrid} & 10 & - & 94.55 \\
 & FCMFS \cite{ma2020filter} & 4.47 & 10.6 & 94.02 \\\hline
\end{tabular}
\label{tab1}
\end{center}
\vspace{-5mm}
\end{table}

\begin{figure}
\vspace{-10mm}
\begin{minipage}[t]{1.05\columnwidth}
  \includegraphics[width=\linewidth]{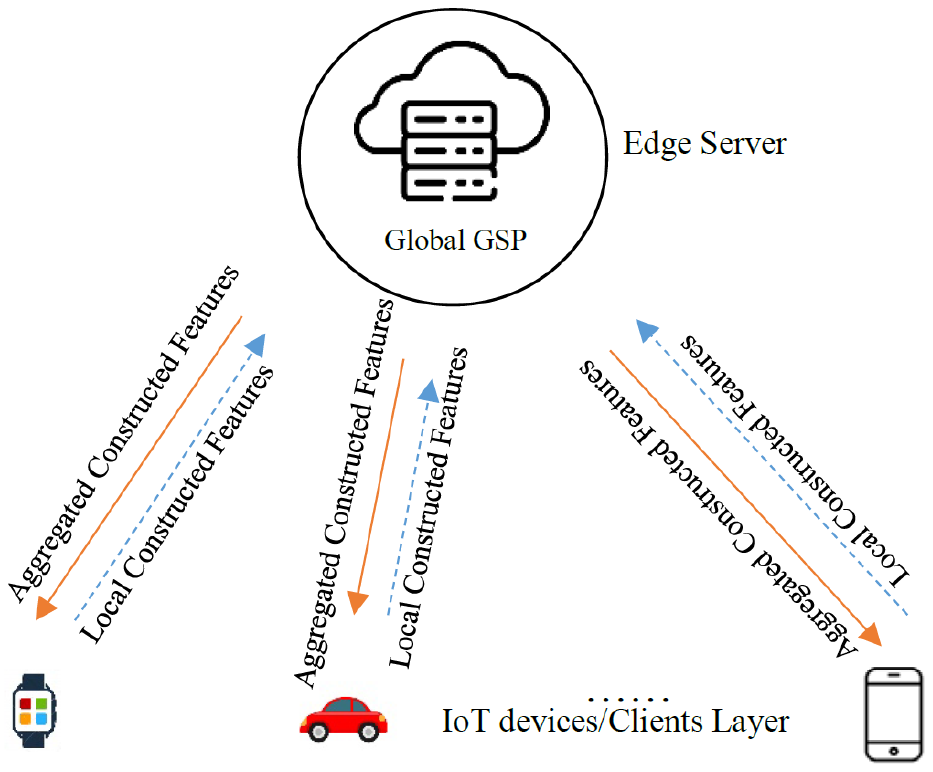}
\end{minipage}\hfill 
\vspace{-58mm}
\caption{Federated Feature Construction Procedure}
\end{figure}

\vspace{-2mm}
\section{Conclusion and Future Works}

In IoT environments, there are many end-user devices that collect enormous amount of data. These datasets may contain some non-informative, noisy, and redundant features. Therefore, data pre-processing methods like feature construction can be effective to reduce data size and learning models' computational costs. In this paper, a multiple federated feature construction method is proposed for the first time. Multimodal optimization with filter-based gravitational search programming are used to create new features in a distributed procedure. To examine the proposed method, three IoT benchmark datasets are fed to a privacy preserving learning method like FL. The results indicate that the proposed method can obtain better accuracy with fewer features compare to other FFS methods. Moreover, the results on eight UCI datasets illustrate that centralized learning models like C4.5 on edge server can obtain accurate results and achieve good trade-off between accuracy and communication cost. In this work, we use crowding clustering strategy for MO and combine it with FFC for the first time. Exploring alternative MO strategies such as speciation and fitness sharing, in combination with FFS and FFC, holds the potential for achieving improved results. Therefore, for future research, we intend to investigate the effectiveness of these methods in distributed pre-processing approaches specifically designed for IoT environments.

\bibliographystyle{IEEEtranN}
\bibliography{References}

\begin{thebibliography}{15}
\providecommand{\natexlab}[1]{#1}
\providecommand{\url}[1]{#1}
\csname url@samestyle\endcsname
\providecommand{\newblock}{\relax}
\providecommand{\bibinfo}[2]{#2}
\providecommand{\BIBentrySTDinterwordspacing}{\spaceskip=0pt\relax}
\providecommand{\BIBentryALTinterwordstretchfactor}{4}
\providecommand{\BIBentryALTinterwordspacing}{\spaceskip=\fontdimen2\font plus
\BIBentryALTinterwordstretchfactor\fontdimen3\font minus
  \fontdimen4\font\relax}
\providecommand{\BIBforeignlanguage}[2]{{%
\expandafter\ifx\csname l@#1\endcsname\relax
\typeout{** WARNING: IEEEtranN.bst: No hyphenation pattern has been}%
\typeout{** loaded for the language `#1'. Using the pattern for}%
\typeout{** the default language instead.}%
\else
\language=\csname l@#1\endcsname
\fi
#2}}
\providecommand{\BIBdecl}{\relax}
\BIBdecl

\bibitem[Nishio and Yonetani(2019)]{nishio2019client}
T.~Nishio and R.~Yonetani, ``Client selection for federated learning with
  heterogeneous resources in mobile edge,'' in \emph{ICC 2019-2019 IEEE
  international conference on communications (ICC)}.\hskip 1em plus 0.5em minus
  0.4em\relax IEEE, 2019, pp. 1--7.

\bibitem[Zebari et~al.(2020)Zebari, Abdulazeez, Zeebaree, Zebari, and
  Saeed]{zebari2020comprehensive}
R.~Zebari, A.~Abdulazeez, D.~Zeebaree, D.~Zebari, and J.~Saeed, ``A
  comprehensive review of dimensionality reduction techniques for feature
  selection and feature extraction,'' \emph{J. Appl. Sci. Technol. Trends},
  vol.~1, no.~2, pp. 56--70, 2020.

\bibitem[Tran et~al.(2019)Tran, Xue, and Zhang]{tran2019genetic}
B.~Tran, B.~Xue, and M.~Zhang, ``Genetic programming for multiple-feature
  construction on high-dimensional classification,'' \emph{Pattern
  Recognition}, vol.~93, 2019.

\bibitem[Yang et~al.(2016)Yang, Chen, Yu, Gu, Li, Zhang, and
  Zhang]{yang2016adaptive}
Q.~Yang, W.-N. Chen, Z.~Yu, T.~Gu, Y.~Li, H.~Zhang, and J.~Zhang, ``Adaptive
  multimodal continuous ant colony optimization,'' \emph{IEEE Transactions on
  Evolutionary Computation}, vol.~21, no.~2, pp. 191--205, 2016.

\bibitem[Cassar{\'a} et~al.(2022)Cassar{\'a}, Gotta, and
  Valerio]{cassara2022federated}
P.~Cassar{\'a}, A.~Gotta, and L.~Valerio, ``Federated feature selection for
  cyber-physical systems of systems,'' \emph{IEEE Transactions on Vehicular
  Technology}, vol.~71, no.~9, pp. 9937--9950, 2022.

\bibitem[Feng(2022)]{feng2022vertical}
S.~Feng, ``Vertical federated learning-based feature selection with
  non-overlapping sample utilization,'' \emph{Expert Systems with
  Applications}, vol. 208, p. 118097, 2022.

\bibitem[Ma and Gao(2020)]{ma2020filter}
J.~Ma and X.~Gao, ``A filter-based feature construction and feature selection
  approach for classification using genetic programming,''
  \emph{Knowledge-Based Systems}, vol. 196, p. 105806, 2020.

\bibitem[Koza(1994)]{koza1994genetic}
J.~R. Koza, ``Genetic programming as a means for programming computers by
  natural selection,'' \emph{Statistics and computing}, vol.~4, pp. 87--112,
  1994.

\bibitem[Li et~al.(2023)Li, Peng, Zhang, Huang, Guo, Yu, and Liu]{li2023fedsdg}
A.~Li, H.~Peng, L.~Zhang, J.~Huang, Q.~Guo, H.~Yu, and Y.~Liu, ``Fedsdg-fs:
  Efficient and secure feature selection for vertical federated learning,''
  \emph{arXiv preprint arXiv:2302.10417}, 2023.

\bibitem[Hu et~al.(2022)Hu, Zhang, Gong, and Sun]{hu2022multi}
Y.~Hu, Y.~Zhang, D.~Gong, and X.~Sun, ``Multi-participant federated feature
  selection algorithm with particle swarm optimizaiton for imbalanced data
  under privacy protection,'' \emph{IEEE Transactions on Artificial
  Intelligence}, 2022.

\bibitem[Mahanipour and Khamfroush(2023)]{mahanipour2023wrapper}
A.~Mahanipour and H.~Khamfroush, ``Wrapper-based federated feature selection
  for iot environments,'' in \emph{2023 International Conference on Computing,
  Networking and Communications (ICNC)}.\hskip 1em plus 0.5em minus 0.4em\relax
  IEEE, 2023, pp. 214--219.

\bibitem[Zhang et~al.(2023)Zhang, Mavromatics, Vafeas, Nejabati, and
  Simeonidou]{zhang2023federated}
X.~Zhang, A.~Mavromatics, A.~Vafeas, R.~Nejabati, and D.~Simeonidou,
  ``Federated feature selection for horizontal federated learning in iot
  networks,'' \emph{IEEE Internet of Things Journal}, 2023.

\bibitem[Mahanipour and Nezamabadi-Pour(2019)]{mahanipour2019gsp}
A.~Mahanipour and H.~Nezamabadi-Pour, ``Gsp: an automatic programming technique
  with gravitational search algorithm,'' \emph{Applied Intelligence}, vol.~49,
  2019.

\bibitem[Hu et~al.(2021)Hu, Zhang, Li, and Deng]{hu2021multimodal}
X.-M. Hu, S.-R. Zhang, M.~Li, and J.~D. Deng, ``Multimodal particle swarm
  optimization for feature selection,'' \emph{Applied Soft Computing}, vol.
  113, p. 107887, 2021.

\bibitem[Ma and Teng(2019)]{ma2019hybrid}
J.~Ma and G.~Teng, ``A hybrid multiple feature construction approach for
  classification using genetic programming,'' \emph{Applied Soft Computing},
  vol.~80, 2019.

\end{thebibliography}
\end{document}